\documentclass[a4paper]{article}

\usepackage{18lomcon}        
\usepackage{cite}             
\usepackage{epsfig}           
\usepackage{epstopdf}

\def\be{\begin{eqnarray}}
\def\ee{\end{eqnarray}}

\begin{document}


\title{PROBLEMS OF SPONTANEOUS AND GRAVITATIONAL BARYOGENESIS}

\author{Elena Arbuzova \email{arbuzova@uni-dubna.ru}
        and
        Alexander Dolgov \email{dolgov@fe.infn.it}
}

\affiliation{$^{a,b}$Novosibirsk State University, Pirogova ul., 2, 630090, Novosibirsk, Russia
}
\affiliation{$^a$Dubna State University, Universitetskaya ul., 19, 141983, Dubna, Russia
}
\affiliation{$^b$ITEP, Bol. Cheremushkinsaya ul., 25, 117259, Moscow, Russia
}

\date{}
\maketitle


\begin{abstract}
Spontaneous and closely related to it gravitational baryogenesis are critically analyzed. It is shown that the coupling of the curvature scalar to baryonic current, 
which induces nonzero baryonic asymmetry of the universe, simultaneously leads to higher order gravitational equations, which have exponentially unstable
 solutions. It is shown that this instability endangers the standard  cosmology.
\end{abstract}

Different scenarios of baryogenesis are based, as a rule, on three well known Sakharov 
principles~\cite{ads}: a) non-conservation of baryonic number; b) breaking of symmetry between particles and antiparticles; c) deviation from thermal equilibrium. For details see e.g. reviews~\cite{AD-BG}. However, none of these conditions is obligatory. 
Only one from three Sakharov principles, namely, non-conservation of baryons, is necessary for 
spontaneous baryogenesis (SBG) in its classical version~\cite{spont-BG},
but this mechanism does not demand an explicit C and CP violation.  It can proceed in thermal equilibrium and it is usually most 
efficient in thermal equilibrium. 

The term "spontaneous" is related to spontaneous breaking of  a global $U(1)$-symmetry, which ensures the 
conservation of the total baryonic number in the unbroken phase. 
When the symmetry is broken, the baryonic
current becomes non-conserved and the Lagrangian density  acquires the term
\be
{\cal L}_{SBG} =  (\partial_{\mu} \theta) J^{\mu}_B\, ,
 \label{L-SB}
 \ee
  $\theta$ is the (pseudo)Goldstone 
  field  and $J^{\mu}_B$ is the baryonic current of matter fields. 

For a spatially homogeneous field $\theta = \theta (t)$  the Lagrangian (\ref{L-SB})
is reduced to ${\cal L}_{SB} =   \dot \theta\, n_B$, where $n_B\equiv J^0_B$ 
is the baryonic number density of matter, so it is tempting to identify $(-\dot \theta)$ with the 
baryonic chemical potential, $\mu_B$, of the corresponding 
system. As is argued in 
refs.~\cite{ad-kf,ADN}, such identification 
is questionable and depends upon the representation chosen for the fermionic fields, but still the scenario is operative and presents a beautiful possibility to create an excess of particles over antiparticles in the universe.

Subsequently the idea of gravitational baryogenesis (GBG) was put forward~\cite{GBG-1}, where the 
scenario of SBG was modified by the introduction of the coupling of the baryonic current to the derivative 
of the curvature scalar $R$:
\begin{eqnarray}
{\cal L}_{GBG} = \frac{f}{m_0^2} (\partial_\mu  R ) J^\mu_B\, ,
\label{L-GBG}
\end{eqnarray}
where $m_0$ is a constant parameter with dimension of mass and $f$ is dimensionless
coupling constant which is introduced for the arbitrariness of the sign.

GBG scenarios possess the same interesting and nice features of SBG, namely generation
of cosmological asymmetry in thermal equilibrium without necessity  of explicit
C or CP violation in particle physics. However, an introduction of the derivative of
the curvature scalar into the Lagrangian of the theory results in high order gravitational
equations which are strongly unstable.  The effects of this instability may drastically 
distort not only the usual cosmological history, but also the standard Newtonian gravitational dynamics. We discovered such instability for  scalar baryons~\cite{ea-ad-gbg-scal}  and  found similar effect for the more usual spin one-half baryons (quarks)~\cite{ea-ad-gbg-ferm}.

Let us start from the model where baryonic number is carried by scalar field $\phi$ with 
potential $U(\phi, \phi^* )$.
The action of the scalar model has the form:
\begin{eqnarray}
A = \int d^4 x\, \sqrt{-g} \left[ \frac{m_{Pl}^2}{16\pi } R + \frac{1}{m_0^2} (\partial_{\mu} R) J^{\mu}  - 
g^{\mu \nu} \partial_{\mu}\phi\, \partial_{\nu}\phi^* + U(\phi, \phi^*)\right] - A_m\, ,
\label{act-tot}
\end{eqnarray}
where $m_{Pl}=1.22\cdot 10^{19}$ GeV is the Planck mass, 
$A_m$ is the matter action,  $J^\mu = g^{\mu\nu}J_\nu$, and $g^{\mu\nu}$ is the metric tensor of the 
background space-time. We assume that initially the metric has the usual GR form and study the emergence 
of the corrections due to the  instability described below.

In the homogeneous case the  equation for the curvature scalar in the FRW metric takes the form:
\be
\frac{m_{Pl}^2}{16\pi }\, R + 
\frac{q^2}{6 m_0^4} \left(R + 3 \partial_t^2 + 9 H \partial_t \right)
\left[ 
\left(\ddot R + 3H  \dot R\right) T^2 \right] + 
\frac{1}{m_0^2} \dot R \, \langle J^0 \rangle 
= - \frac{T^{(tot)}}{2},
\label{trace-eq-plasma}
\ee
where  $\langle J^0 \rangle $ is the thermal average value of the baryonic number density of $\phi$,
$q$ is the baryonic number of $\phi $,
$H = \dot a/a$ is the Hubble parameter, and $T^{(tot)}$ is the trace of the energy-momentum tensor of matter including 
contribution from the $\phi$-field. In the homogeneous and isotropic cosmological plasma 
\be
T^{(tot)} = \rho - 3 P\, ,
\label{T-tot}
\ee
where $\rho$ and $P$ are respectively the energy density and the pressure of the plasma. 
For relativistic plasma $\rho = \pi^2 g_* T^4/30$ with $T$ and $g_*$ 
being the plasma temperature and the number of particle species in the plasma. 
The Hubble parameter is expressed through $\rho$ as $H^2 = 8\pi \rho/(3m_{Pl}^2) \sim T^4/m_{Pl}^2$.

Keeping only the linear in $R$ terms and neglecting higher powers of $R$, such as $R^2$ or $H R$, we obtain the
linear fourth order differential equation:
\be
\frac{d^4 R}{dt^4} + \mu^4 R =  - \frac{1}{2} \, T^{(tot)}\,,\  {\rm where }\ \   \mu^4 = \frac{m_{Pl}^2 m_0^4}{8 \pi q^2 T^2}\,.
\label{d4-R}
\ee

The homogeneous part of this equation has exponential solutions  $R \sim \exp (\lambda t)$ with $\lambda = | \mu | \exp \left(  i\pi /4 + i \pi n /2 \right)$, where $n = 0,1,2,3$. There are two solutions with positive real parts of $\lambda$.
This indicates that the curvature scalar is
exponentially unstable with respect to small perturbations, so $R$ should rise exponentially fast with time 
and quickly oscillate around this rising function.

Now we need to check if the characteristic rate of  the perturbation explosion is indeed much larger than the rate
of the universe expansion, that is:
\be 
(Re\, \lambda)^4   > H^4 = \left( \frac{ 8\pi \rho }{ 3 m^2_{Pl}}\right)^2 =  \frac{16 \pi^6 g_*^2}{2025}\,
\frac{ T^8}{m_{Pl}^4}, 
\label{lambda-to-H}
\ee
where $\rho = \pi^2 g_* T^4 /30$ is the energy density of the primeval plasma at temperature $T$
and $g_* \sim 10 - 100$ is the number of relativistic
degrees of freedom in the plasma. This condition is fulfilled if
\be
    \frac{2025}{2^9 \pi^7 q^2 g_*^2}\frac{m_{Pl}^6 m_0^4}{T^{10}} > 1\,,
\label{inst-OK}
\ee
or, roughly speaking, if $T \leq m_{Pl}^{3/5} m_0^{2/5} $.   At these temperatures the
instability is quickly developed and the standard cosmology would be destroyed.

Let us now generalize results, obtained for scalar baryons, to realistic fermions. We start from the action in the form
\begin{eqnarray}  \nonumber
A&=& \int d^4x \sqrt{-g} \left[\frac{m_{Pl}^2}{16 \pi} \,R - {\cal L}_{m}\right]\, , \ \   \\  \nonumber
\label{A-gen}
{\cal L}_{m} &= &\frac{i}{2} (\bar Q \gamma ^\mu \nabla _\mu Q -   \nabla _\mu \bar Q\, \gamma ^\mu Q) - m_Q\bar Q\,Q\\ \nonumber
&+& 
\frac{i}{2} (\bar L \gamma ^\mu \nabla _\mu L -   \nabla _\mu \bar L \gamma ^\mu L)
- m_L\bar L\,L \\ 
&+& \frac{g}{m_X^2}\left[(\bar Q\,Q^c)(\bar Q L) + (\bar Q^cQ)(\bar L Q) \right]
+ \frac{f}{m_0^2} (\partial_{\mu} R) J^{\mu} + {\cal L}_{other}\,, 
\end{eqnarray}
where $Q$ is the quark (or quark-like) field with non-zero baryonic number, $L$ is
another fermionic field (lepton),
$\nabla_\mu  $ is the covariant derivative of Dirac fermion in tetrad formalism, 
$J^{\mu } = \bar Q \gamma ^{\mu} Q$ is the quark current with $\gamma ^{\mu}$ being the curved space gamma-matrices,  
${\cal L}_{other}$ describes all other forms of matter.
 \nopagebreak The four-fermion interaction between quarks and leptons is introduced to ensure the
necessary non-conservation of the baryon number with $m_X$ being a constant
parameter with dimension of mass and $g$ being a dimensionless coupling constant.
In grand unified theories $m_X$ may be of the order of $10^{14}-10^{15}$ GeV.

Varying  the action (\ref{A-gen}) over metric, $g^{\mu \nu}$, and taking trace with respect to $\mu$ and $\nu$, we obtain the following equation of motion for the curvature scalar:
\begin{eqnarray} \nonumber
- \frac{m_{Pl}^2}{8\pi } R &=& m_{Q} \bar Q Q + m_L \bar L L 
+
 \frac{2g}{m_X^2}\left[(\bar Q\,Q^c)(\bar Q L) + (\bar Q^cQ)(\bar L Q) \right] \\
 &-& 
 \frac{2f}{m_0^2} (R + 3D^2) D_{\alpha} J^{\alpha} + T_{other}\,,
 \label{trace}
 \end{eqnarray}
where $T_{other} $ is the trace of the energy momentum tensor of all other fields.
At relativistic stage, when masses are negligible, we can take  $T_{other} = 0$. 
The average  expectation value of the interaction term proportional to $g$ is also 
small, so the contribution of all matter fields may be neglected.

We used the kinetic equation, which leads to an explicit dependence on $R$
of the current divergence, $D_\alpha J^\alpha$, if the current is not conserved. As
a result we obtain 4th order differential equation for $R$:
\begin{eqnarray}
\frac{d^4 R}{dt^4} = \lambda^4 R,\  {\rm where }\ \ \lambda^4 = \frac{5 m_{Pl}^2 m_0^4} {36\pi  g_s B_q f^2 T^2}.
\label{d4-R}
\end{eqnarray}
Here $g_s$ and $B_q$ are respectively
 the number of the spin states and the baryonic number of quarks.  Deriving this equation we neglected
the Hubble parameter factor in comparison with time derivatives of $R$.

Evidently equation (\ref{d4-R}) has extremely unstable solution with instability time
by far shorter than the cosmological time. This instability would lead to an explosive
rise of $R$, which may possibly be terminated  by the nonlinear  terms proportional to the
product of $H$ to lower derivatives of $R$. Correspondingly one may expect stabilization when $HR \sim \dot R$,
i.e. $H\sim \lambda$. Since $\dot H + 2 H^2 = - R/6$,
$H$ would also exponentially  rise together with  $R$, 
$ H \sim \exp (\lambda t )$ and $\lambda H \sim R$.
Thus stabilization may take place at $R \sim \lambda^2   \sim  m_{Pl} m_0^2 / T$.
This result  should be compared with the normal General Relativity value
$ R_{GR} \sim T_{matter} /m_{Pl}^2 $, where $T_{matter}$ is the trace of the 
energy-momentum tensor of matter.

\section*{Acknowledgments}
This work  was supported by the RSF Grant N 16-12-10037. 

\end{document}